\documentstyle[psfig]{mn}

\newcommand{\rmn}[1]{\mathrm {#1}}

\pagerange{\pageref{firstpage}--\pageref{lastpage}}

\pubyear{1996}

\begin{document}

\title[Hot gas recycling in ellipticals]{A homologous
recycling model for hot galactic coronae} 
\author[A. G. Kritsuk]{Alexei G. Kritsuk$^{1,2,3}$\thanks{E-mail: \tt
agk@aispbu.spb.su}\\
$^1$ Institute of Astronomy, University of St.
Petersburg, Stary Peterhof, St. Petersburg 198904, Russia\\ 
$^2$ Max-Planck-Institut f\"ur Astrophysik, Postfach
1523, D-85740 Garching, Germany\\
$^3$Max-Planck-Institut f\"ur extraterrestrische Physik, 
Postfach 1603, D-85740 Garching, Germany}

\date{Accepted 1995 December 11.
      Received 1995 July 6;
      in original form 1995 January 3}

\maketitle

\label{firstpage}

\begin{abstract}
An equilibrium model is presented for a hydrostatic
isothermal hot gas distribution in a gravitational well of a giant
elliptical galaxy immersed in a massive dark halo.
The self-consistently determined gravitational potential of the system
provides the optical surface brightness distribution of the galaxy,
matching the de Vaucouleurs $R^{1/4}$-law, if the sound speed in the gas
$c$, stellar velocity dispersion $\sigma_*$, and velocity dispersion
of dark matter $\sigma_{\rmn{DM}}$ are related by
$\sigma_{\rmn{DM}}\simeq c\simeq\sqrt{2}\sigma_*$. 
While the first equality follows from virial considerations, 
the second one relies on the observed similarity of optical and X-ray
brightness profiles. The thermal equilibrium of
the gas is described in the framework of a one-zone model, which
incorporates radiative cooling, stellar mass loss, supernova heating,
and mass sink due to local condensational instability. The sequence of
{\em saddle-node} and {\em saddle-connection} bifurcations provides a
first-order phase transition, which separates the stable hot phase in the
cooling medium for a reasonably high efficiency of condensation and
precludes catastrophic cooling of the gas. The
bifurcations occur naturally, due to the shape of the radiative cooling
function, and this holds for a wide range of gas metallicities. The one-zone
model implies a scaling relation for the equilibrium stellar and gas densities 
$\rho_*\propto\rho^2$, which allows a
hydrostatic, thermally stable distribution for the hot isothermal
coronal gas. The resulting homologous recycling model reproduces the basic
optical and X-ray properties of relatively isolated giant elliptical
galaxies with hot haloes.
Implications for cD galaxies in X-ray dominant clusters and groups are
briefly discussed in the general context of galaxy formation and evolution. 
\end{abstract}


\section{Introduction} The analysis of the extended X-ray emission of 
early-type galaxies has led to an extension of the cooling flow model [see
Fabian, Nulsen \& Canizares \shortcite{fabian..84} and Fabian 
\shortcite{fabian94}] 
applied to the class of bright elliptical galaxies 
\cite{thomas....86}. The typical X-ray luminosities of such galaxies
are in the range $10^{39}-10^{42}$ erg s$^{-1}$, the gas
temperature is $5\times10^6-2\times10^7$~K, and the electron number
density in the centre is $0.01-0.1$ cm$^{-3}$. The galaxies usually contain
$10^9-10^{10}$M$_{\odot}$ of the gas in their extended hot haloes
\cite{forman..85}. 
Observations reveal a tight connection of X-ray and optical
characteristics of the galaxies, including a strong correlation
between the X-ray and optical luminosities, although with a considerable
scatter \cite{canizares..87,donnelly..90}, and a similarity of X-ray and 
optical light distributions (Trinchieri, Fabbiano \& Canizares 1986; Killeen
\& Bicknell 1988). \nocite{trinchieri..86} \nocite{killeen.88}

The source of the gas in hot haloes of early-type galaxies is
usually attributed to the normal stellar mass loss with a specific rate
$\dot\rho_*/\rho_*\sim1-2\times10^{-12}$ yr$^{-1}$. Gas injection from
stars and supernova events in the galaxy make the gas inhomogeneous
\cite{mathews90} and thermal instability causes matter to cool out at
radii $r<r_{\rmn{cool}}$, where $t_{\rmn{cool}}<H_0^{-1}$ \cite{nulsen86}. The
dynamics of the gas, therefore, is determined by the relative rates of
the processes, controlling its mass and energy balance. Theoretical
work done so far has demonstrated a variety of possible behaviours for
the gas, ranging from supersonic galactic winds to subsonic inflows [see
Thomas \shortcite{thomas86}, Hattori, Habe \& Ikeuchi \shortcite{hattori..87},
David, Forman \& Jones \shortcite{david..90}, Ciotti et al. 
\shortcite{ciotti...91} 
and references therein]. However, due to the  
non-linear nature of radiative cooling, numerical modelling of inflows
usually ends up with a cooling catastrophe in the centre of the
galaxy, which overproduces the observed X-rays and precludes the study
of further evolution. As there are no certain observational
indications of such a catastrophe, it is important to understand
the nature of possible physical damping mechanisms, which are able to
prevent it. 

It is an objective of this paper to present a purely hydrostatic,
thermodynamically stable model for gas recycling in an elliptical
galaxy, which satisfies the observational constraints. The model
incorporates all the same physical processes as the `standard' multiphase
cooling flow model, and may also be useful in interpreting the X-ray
emission from the central `cooling flow' regions in dynamically quiet
X-ray dominant clusters or groups, where the outskirts of the central 
dominant optical galaxy extend to the cooling radius.  

\section{Underlying galaxy model}\label{GalMod}
Let $M(r)$ be a spherically symmetric configuration
of the gravitating mass, including the stellar component of a galaxy, 
the hot gas,
and the dark halo. Then the hydrostatic gas distribution in the gravitational
potential $\phi(r)$ of $M(r)$ can be described by
\begin{equation}
{kT\over\mu m_{\rmn{p}}}\left(\frac{\rmn{d}\;
\ln{\rho}}{\rmn{d}\:{r}} + \frac{\rmn{d}\;
\ln{T}}{\rmn{d}\:{r}}\right)=-\nabla\phi\equiv -{GM(r)\over r^2}. 
\end{equation}
For the isothermal gas this reduces to
\begin{equation}
{kT\over\mu m_{\rmn{p}}}\;\frac{\rmn{d}\;
\ln{\rho}}{\rmn{d}\:{r}}=-\nabla\phi. \label{IsothHydG} 
\end{equation}
Here $T$ and $\rho$ are the gas temperature and density, 
$\mu$ the mean molecular weight, 
$m_{\rmn{p}}$ the proton mass, and $k$ Boltzmann's constant.

The stellar density distribution of the galaxy, $\rho_*$, responding to the
same gravitational potential, satisfies
\begin{equation}
\sigma_{\rmn{r}}^2\left(\frac{\rmn{d}\: 
\ln{\rho_*}}{\rmn{d}\:{r}}+\frac{\rmn{d}\;
\ln{\sigma^2_{\rmn{r}}}}{\rmn{d}\:{r}}+{2\eta\over
r}\right)=-\nabla\phi,
\end{equation}
where $\sigma_{\rmn{r}}$ is the radial velocity dispersion of stars in the
galaxy, the anisotropy parameter $\eta$ is defined as 
$\eta=1-\sigma^2_{\rmn{t}}/\sigma_{\rmn{r}}^2$, and the subscript `t' 
represents 
the tangential velocity component. In the case of an isotropic and
uniform stellar velocity distribution 
$(\sigma_{\rmn{t}}=\sigma_{\rmn{r}}\equiv const)$ 
it simply follows that 
\begin{equation}
\sigma_*^2\frac{\rmn{d}\: \ln{\rho_*}}{\rmn{d}\:{r}}=-\nabla\phi, 
\label{IsothHydSt}
\end{equation}
where $\sigma_*$ is the one-dimensional velocity dispersion.
For an isothermal gas and an isotropic, constant stellar
velocity dispersion the stellar and gas densities are related as
\begin{equation}
\frac{\rmn{d}\: \ln{\rho}}{\rmn{d}\:\ln{\rho_*}}={\mu m_{\rmn{p}} 
\sigma^2_*\over kT}
={\sigma^2_*\over c^2}\equiv\beta,\label{Beta}
\end{equation}
or
\begin{equation}
\rho=C\rho_*^{\beta},\label{GasDenPowerLaw}
\end{equation}
where $c$ is the isothermal sound speed and $C$ is the integration
constant. This relation provides the basis for the so-called isothermal  
$\beta$-model \cite{cavaliere.76}, extensively used to describe the
X-ray surface brightness of the hot gas in clusters of galaxies.

Given equation (\ref{GasDenPowerLaw}) and assuming similar
profiles for the gas density $\rho$ and for the dark matter density
$\rho_{\rmn{DM}}$ (and, therefore, $\sigma_{\rmn{DM}}=c$)\footnote{The 
approximate
equality of the dark matter velocity dispersion and the sound velocity of
the gas has been demonstrated, e.g. by the simulations of X-ray clusters of
galaxies in the cold dark matter cosmogony for those clusters that
are not merging objects, experiencing a transient boost in the
velocity dispersion of the system \cite{Navarro..94}.}, one can define
the total mass distribution self-consistently, using Poisson's
equation   
\begin{equation}
{1\over r^2}{\rmn{d}\over \rmn{d}r}\left(r^2{\rmn{d}\ln\rho_*\over
\rmn{d}r}\right)=-{4\pi 
G\over \sigma_*^2}(\rho_*+\rho+\rho_{\rmn{DM}}). \label{Poi}
\end{equation}
When $\beta\simeq1/2$ and in the centre $\rho_{*,0}$ is somewhat
higher than $\rho_{\rmn{DM},0}$, the stellar system has a density distribution
which is similar to the profile of a self-gravitating isothermal
sphere, but has a density cut-off at the core radius of the dark matter
halo, where the dark matter density begins to fall as
$\rho_{\rmn{DM}}\propto r^{-2}$, cf. Burkert \shortcite{burkert94}. 
It turns out 
that such a solution of (\ref{Poi}) gives the projected stellar mass
density, which follows the de Vaucouleurs $R^{1/4}$-law over several
orders of magnitude in surface brightness, in a wide range of radii
$0.1R_{\rmn{e}}\le r\le1.5R_{\rmn{e}}$, where $R_{\rmn{e}}$ is the  
effective radius \cite{burkert94}. This can be expected if 
the stellar system formed in a dissipative process from 
post-infall shock heated gas with the star formation rate
$\propto\rho^2$. 

The volume density of the $\beta=1/2$ stellar system falls more steeply
than $\rho_*\propto r^{-2}$ outside the core region \cite{burkert94}, 
therefore the 
restriction on stellar velocity dispersion, imposed to derive
(\ref{GasDenPowerLaw}), can be formulated as
\begin{equation}
\left|{\rmn{d}\; \ln{\sigma_*}\over
\rmn{d}\:\ln{r}}+\eta\right|\ll 1,
\end{equation}
while the gas must be isothermal up to 
\begin{equation}
\left|{\rmn{d}\; \ln{T}\over
\rmn{d}\:\ln{r}}\right|\ll 1.
\end{equation}
Due to projection effects it is difficult to demonstrate
whether the above conditions are indeed satisfied in observed systems
or not. However, in most cases early-type galaxies do exhibit rather flat
line-of-sight velocity dispersion profiles and observations also show that
the gas is quite isothermal. 
The details of observational data will be briefly discussed
in Section \ref{Obs}. 

\section{One-zone model revisited} A simple semi-analytical one-zone
model allows an insight into the non-linear
thermodynamics of the hot gas in the halo of a galaxy. While the stability
of gas equilibria was studied in Kritsuk \shortcite{kritsuk92} with an 
emphasis on
limit cycle solutions, considerations here concentrate mainly on
classification of bifurcations, which lead to the separation of a stable hot
phase of the thermally unstable cooling medium. In the following
subsections the basic formulae of the model are compiled from Kritsuk  
\shortcite{kritsuk92} in order to lead up to the bifurcation analysis. The
only new physical process introduced is the feedback heating. It
was neglected before, but may play an important role in highly
non-equilibrium situations.
\subsection{Input physics} The set of physical processes
incorporated in cooling flow models includes radiative cooling, mass
deposition due to thermal instabilities, supernova (SN) type Ia heating,
and stellar mass loss for the galaxy, which is immersed in a
dark halo. Here only the basic formulae are
presented; one can find more extensive discussion elsewhere 
[e.g. in Sarazin \& White \shortcite{sarazin.87}, see also references
below].
The standard notation is kept where possible.
\subsubsection{Stellar mass loss}
The efficiency of gas deposition due to stellar mass loss and SN Ia
explosions is proportional to the stellar mass density: 
$\alpha \rho_*(r)$, where $\alpha=\alpha_*+\alpha_{\rmn{sn}}\equiv const$
incorporates contributions from more quiescent stellar mass loss
$\alpha_*\simeq 4.7\times10^{-20}$ s$^{-1}$
\cite{faber.76b,sarazin.87}, and mass loss by supernovae
$\alpha_{\rmn{sn}}=1.33\times10^{-21}$ s$^{-1}$ (the estimate
corresponds to $r_{\rmn{sn}}=0.24\:h_{50}^2$
SNu, $M_{\rmn{sn}}=1.4$ M$_{\odot}$, and $(M/L_B)_*=8$M$_{\odot}/$L$_{\odot}$).
The supernova type Ia rate, $r_{\rmn{sn}}$, is uncertain by a
factor of the order of 1.5 according to van den Bergh \& Tammann 
\shortcite{vandenbergh.91}, where
$h_{50}=H_0/50$ km s$^{-1}$Mpc$^{-1}$, but a significantly lower estimate
$r_{\rmn{sn}}=0.06\pm0.03\:h_{50}^2$ SNu is given by Turatto, Cappellaro \&
Benetti \shortcite{turatto..94}.
\subsubsection{Thermal instabilities} Drops of condensed material cool
out from the hot gas because of thermal instabilities at the rate of 
$\dot\rho_{\rmn{ti}}=b\chi(n)\rho$, where the dimensionless parameter 
$b\equiv const\in [0,\; 1]$ controls the efficiency of condensation. 
The Heaviside
function $\chi$ and the instability growth rate $n$ are defined as	
\begin{equation}
\chi(n)=\cases{n, &if $n\ge 0$;\cr0, &otherwise;} 
\end{equation}
\begin{equation}
n\equiv {\partial \over \partial t} 
\left(\ln{\delta \rho \over \rho}\right)={1\over
c_p}\left({2\rho\Lambda\over T}-\rho{\rmn{d}\Lambda\over \rmn{d} T}\right) -
{\alpha\rho_*\over \rho}. \label{GrowthRate}
\end{equation}
Here $c_p$ is the specific heat at constant pressure, the first term in
the rhs of equation (\ref{GrowthRate}) coincides with Field's
instability criterion, and the second describes stabilization due to stellar
mass loss \cite{kritsuk92}. Note that if $\rho_*\propto\rho^2$,
as in the 
$\beta=1/2$ case, then $\dot \rho_{\rmn{ti}}\propto\rho^2$, which naturally
implies a star formation rate from the  cooled material proportional
to the gas density squared. 
\subsubsection{Radiative cooling} Non-equilibrium radiative cooling as
a function of temperature and metallicity, $\Lambda(T,Z)$ erg
cm$^3$ g$^{-2}$ s$^{-1}$, is compiled from the calculations of Sutherland \& 
Dopita 
\shortcite{sutherland.93}, where solar abundance ratios were taken from
Anders \& Grevesse \shortcite{anders.89}.
Effects of dust on the cooling are not taken into
account here since the coronal phase is believed to be essentially
dust-free. If the gas-to-dust ratio were similar to Galactic values,
and the dust were well mixed in the hot interstellar medium, it could be 
the dominant 
cooling agent at gas temperatures above $2\times10^6$~K \cite{burke.74}. 
This would essentially change the shape of the cooling function
and therefore the gas stability properties outlined below. A crude estimate
shows that dust cooling can be safely ignored if the dust content is $\la0.01$ 
of the Galactic value.
\subsubsection{Heating} The rate of heating due to thermalization of
stellar winds and due to type Ia SN events is assumed to be equal to
$\alpha\rho_*T_0$, where $T_0$ is the characteristic
temperature of the heat source,
$T_0=(\alpha_*T_*+ \alpha_{\rmn{sn}}T_{\rmn{sn}})/\alpha$.
Heating by stellar motions is described by
$T_*=\mu m_{\rmn{p}}\sigma_*^2/k=6.47\times10^6$~K (an estimate is
made for $\sigma_*=300$ km s$^{-1}$), and SN temperature  $T_{\rmn{sn}}={\mu
m_{\rmn{p}}v_{\rmn{ej}}^2 \over3k}=1.09\times10^9$~K
for $E_{\rmn{sn}}=6\times10^{50}$
 erg, $M_{\rmn{sn}}=1.4$ M$_{\odot}$. With these values
$T_0\simeq 3.6\times10^7$~K, while for $r_{\rmn{sn}}=0.06$ SNu the source
temperature is lower: $T_0\simeq 1.7\times10^7$~K.
\subsubsection{Feedback heating} Energy feedback by type II (Ib) SN
explosions due to star formation in cooled material can serve as an
additional source of gas heating. 
This slightly modifies formulae for $\alpha$ and $T_0$: 
$\alpha=\alpha_*+\alpha_{\rmn{sn}}+\alpha_{\rmn{fb}}$, where
$\alpha_{\rmn{fb}}=\varepsilon_{\rmn{fb}}
\dot\rho_{\rmn{ti}}$,
$T_0=(\alpha_*T_*+\alpha_{\rmn{sn}}T_{\rmn{sn}}+\alpha_{\rmn{fb}}
T_{\rmn{fb}})/\alpha$.
Here 
$T_{\rmn{fb}}={2\mu m_{\rmn{p}} E_{\rmn{th}} 
\over3k M_{\rmn{ej}}}=3.6\times10^9$~K,
 $E_{\rmn{th}}=0.72E$ in the Sedov stage, and the supernova explosion energy  
$E=2\times10^{51}$ erg. The optimistic estimate  
$\varepsilon_{\rmn{fb}}=0.01$ is given in Silk et al. 
\shortcite{silk...86}, assuming a
rate of type Ib SNe of 1 supernova yr$^{-1}$ per 100 M$_{\odot}$
yr$^{-1}$, and the mass of ejecta returned to the ISM
$M_{\rmn{ej}}=1$~M$_{\odot}$. The associated mass feedback is
negligible, i.e. $\alpha\gg\varepsilon_{\rmn{fb}}\alpha$.

\subsection{Basic equations} The mass and energy balance of the hot
gas in the halo of a galaxy may be described locally by the autonomous
set of ordinary differential equations:
\begin{equation}
{\rmn{d}\rho\over \rmn{d} t}=\alpha\rho_*-\dot\rho_{\rmn{ti}},\label{Den}
\end{equation}
\begin{equation}
{ \rmn{d} T\over \rmn{d} t}= {1\over\rho}\left[\alpha\rho_*(T_0-T)
-(\gamma-1)\dot\rho_{\rmn{ti}}T - \rho^2\Lambda/c_{\rmn{v}}\right],\label{Tem}
\end{equation}
where $t$ is the time variable and $\alpha$, $\rho_*$, and $T_0$ are
treated as constants; $c_{\rmn{v}}$ is the specific heat at constant
volume. The evolutionary solutions of equations
(\ref{Den}) and (\ref{Tem}) 
may be represented as trajectories in the $(\rho, T)$ phase 
plane. The steady states of the system correspond to the fixed points
and may be readily found as
\begin{equation}
\rho=\rho_{\rmn{eq}}c_0\sqrt{\alpha\rho_*\over 
(\gamma-1)\Lambda_0}, \label{GasDen}
\end{equation}
\begin{equation}
T=\Theta_{\rmn{eq}}T_0,\label{GasTem}
\end{equation}
where $c_0=c(T_0)$ is the isothermal sound speed,
$\Lambda_0=\Lambda(T_0,Z)$, metallicity $Z$ is assumed to be constant,
and $\rho_{\rmn{eq}}$ and $\Theta_{\rmn{eq}}$ 
are dimensionless functions of $b$,
$T_0$, and $Z$: 
\begin{equation}
\left({1\over\gamma\Theta_{\rmn{eq}}}-1\right)(2-\nu_{\rmn{eq}})=1+{1\over b},
\label{Teq} 
\end{equation}
\begin{equation}
\rho_{\rmn{eq}}^2=(1-\gamma\Theta_{\rmn{eq}})/\lambda_{\rmn{eq}}.
\end{equation}
The subscript `eq' denotes an equilibrium state,
$\nu=\rmn{d}\ln{\Lambda}/\rmn{d}\ln{T}$, and $\lambda=\Lambda/\Lambda_0$.

The characteristic time-scale for evolution near the equilibrium,
\begin{equation}
t_{\rmn{s}}=\sqrt{t_{c_0}t_{\alpha}}, \label{TimeScale}
\end{equation}
is a geometric mean of the local cooling time 
$t_{c_0}\equiv c_0^2/[(\gamma-1)\rho\Lambda_0]$ 
and the time-scale for stellar mass loss 
$t_{\alpha}=(\alpha\rho_*/\rho)^{-1}$.

For a given metallicity of the gas $Z$, stellar mass loss rate
$\alpha$, temperature of the heat source $T_0$, and efficiency of
condensation $b$, one can find the equilibrium solutions, if the
cooling function is known. It is a straightforward piece of algebra
then to classify these equilibria according to their stability properties,
using the standard linear technique (see Kritsuk 1992)\nocite{kritsuk92}. This
analysis relies on poorly known high-order derivatives of $\Lambda$
with respect to temperature 
and, therefore, the fine details of stability changes cannot be well
resolved. At the same time, the main stability features, controlled by the
nature of radiative cooling (resonance line emission, producing a
prominent maximum in the range of temperatures 
$10^{4.2}$~K $\la T\la10^{7.2}$~K, 
and thermal bremsstrahlung for $T\ga10^{7.2}$~K), do
provide physically important effects, which are discussed in the next
section.  

\subsection{First-order phase transition in a system with no feedback} 
The efficiency of condensation $b$ is the only essentially free
parameter of the model. It is important to understand how the
equilibria and their stability follow the changes of this control
parameter. 

\begin{figure*}
\psfig{figure=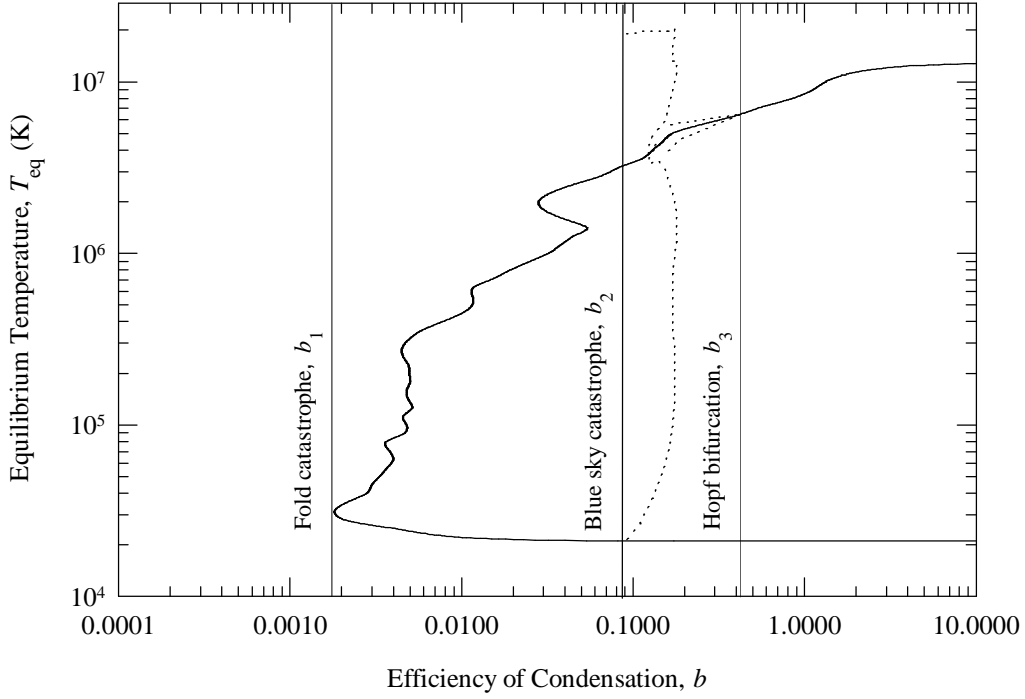,clip=}
\vspace{-17.5cm}
\caption{Bifurcation diagram for $T_0=3\times10^7$~K and $Z=$ Z$_{\odot}$.
The solid line shows the equilibrium temperature $T=\Theta_{\rmn{eq}}T_0$ as
a function of condensation efficiency $b$. Dotted lines indicate 
minimum and maximum temperature values for limit cycle solutions. The
phase transition occurs at $b=b_2$. }
\label{BifDiag}
\end{figure*}

This is illustrated with a plot of the equilibrium temperature
$T_{\rmn{eq}}$ against $b$ for a fixed  value of the heat source temperature
$T_0=3\times10^7$~K and solar metallicity (see
Fig.~\ref{BifDiag}).
There are no steady-state solutions when the condensation efficiency
is low, $b<b_1=0.0018$. The catastrophic cooling is unavoidable even
in a system with reasonably strong SN heating. At $b=b_1$, as
a result of a {\em fold} bifurcation, a {\em saddle-node} pair of fixed
points appears simultaneously, providing stable and unstable
steady-state solutions, classified as a {\em node} (higher temperature state)
and a {\em saddle} (lower temperature state), respectively. The fold
bifurcation is the simplest form of {\em local} bifurcation in that it
requires only a one-dimensional phase space, only one control
parameter, and only lowest-order 
terms (see, e.g., Guckenheimer \& Holmes 1983;\nocite{guckenheimer.83} 
Thompson \& Stewart 1989)\nocite{thompson.89}. It is {\em
structurally stable} and, therefore, the small perturbation of the
dynamics does not alter the bifurcation qualitatively. Equation
(\ref{Teq}) shows that the condensation efficiency $b$ is a unique
function of the equilibrium temperature $\Theta_{\rmn{eq}}$ and depends
solely on the logarithmic temperature derivative of the cooling
function $\nu_{\rmn{eq}}$. Therefore the fold {\em catastrophe}\footnote{The
notion of {\em 
catastrophe} is introduced in non-linear dynamics to specify a
spontaneous birth of an equilibrium state at a critical point.
A fold is an
example of such a {\em discontinuous} bifurcation. Its
normal form, which embodies the local low-order approximation, may be  
written as $\dot x=\mu-x^2$, where for control $\mu>0$ a pair of
equilibria exists, one attracting and the other repelling. For $\mu<0$
the solutions diverge to $x\rightarrow-\infty$ as $t\rightarrow\infty$
\cite{thompson.89}.}
occurs in the model due to the local structure of $\Lambda(T)$. As the 
control parameter becomes larger, multiple steady states appear and
annihilate due to a series of bifurcations of the same kind. These
further local changes in the phase portrait of the system at
$b_1<b<0.054$ are not so important because the new stable solutions
have negligibly small {\em domains of attraction} (or basins) in the
phase plane. 
The most essential event in the range of $b\in[0,\:0.6]$ happens at
$b=b_1$, where the first pair of fixed points spontaneously appears.
Although these changes in the dynamics do not preclude the catastrophic
cooling, the birth of the saddle leads up to further transformations in
the system.

\begin{figure*}
\vspace{-2cm}
\psfig{figure=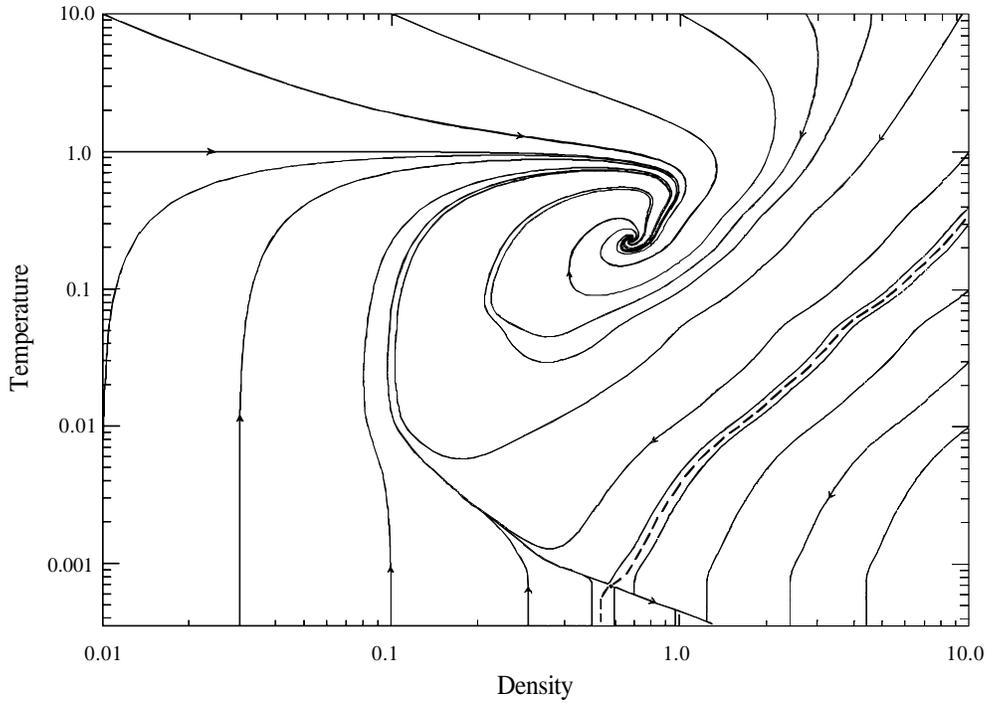,clip=}
\vspace{-17.9cm}
\caption{Phase portrait for one-zone model with $T_0=3\times10^7$~K,
$b=0.5$, and $Z=$ Z$_{\odot}$. Solid lines with arrows show trajectories in
the phase plane 
(temperature and density are dimensionless). Dashed lines indicate the
insets of the {\em saddle} or {\em separatrices}, which separate the
attraction domain of the stable {\em spiral} point.}
\label{Port}
\end{figure*}

The next important event takes place at $b_2=0.092$ due to a
spontaneous birth of a stable limit cycle. This is a {\em homoclinic
connection} bifurcation or a {\em blue sky} catastrophe for a periodic
limit cycle  \cite{guckenheimer.83,thompson.89}. The
blue sky catastrophe is a {\em global} bifurcation\footnote{The
profound changes 
in qualitative behaviour of the system near the threshold value $b_2$
are associated solely with the {\em global} topological configuration of the
{\em insets} and {\em outsets} (the incoming and outcoming critical 
trajectories) 
of the saddle fixed point, which requires a finite
region in the phase plane and cannot be described locally. A saddle
connection bifurcation is known to occur, for example, in the averaged
autonomous van der Pol system [see  equation (2.1.14) of Guckenheimer \& 
Holmes 
\shortcite{guckenheimer.83}].}  
in the phase plane in which the {\em attracting} limit cycle disappears
(as $b$ decreases) by collision with a saddle. The bifurcation is
also structurally stable, that is, the full control-phase space
diagram may be deformed, but qualitatively unchanged by small
perturbations of the governing equations.  At $b>b_2$ the basin of
the attracting limit cycle grows, while the amplitude of non-linear
temperature oscillations becomes smaller. As a result, at $b>0.2$ only
small-amplitude stable oscillations survive, which disappear at
$b=b_3=0.427$ due to a supercritical Hopf bifurcation. Fig. \ref{Port} shows
the huge basin of the stable spiral point at $b=0.5$. This
means that,
whatever the initial temperature and density of the gas are, the
thermodynamic state of the gas settles to an equilibrium with 
temperature $\sim10^7$~K (for $T_0\simeq10^{7.5}$~K) in a finite time
of the order of $t_{\rmn{s}}$ (see equation \ref{TimeScale}). What actually 
happens at the threshold value $b_2$ is 
a first-order phase transition, which separates the hot phase
in the cooling thermally unstable medium. Thus, one has a stable thermodynamic 
description for a gas with embedded cold condensations for
$b\ge b_{\rmn{crit}}\approx0.2$, when the oscillations are
small. The occurrence of
the relaxation oscillations in the vicinity of a saddle-connection
follows from theorem 6.1.1 in Guckenheimer \& Holmes 
\shortcite{guckenheimer.83}.  A detailed
discussion of the cooling function properties, which result in the
saddle connection, can be found in Kritsuk \shortcite{kritsuk92}. 

The key equation (\ref{Teq}) does not contain an absolute value of the
cooling, therefore one can conclude that, although the cooling is less
efficient for metallicities lower than solar, the structure of
equilibrium states, and, in particular, the equilibrium temperature
$T(b,Z)$ does not change dramatically with $Z$. This is actually
the case, as follows from the calculations based on the zero-field
non-equilibrium ionization
cooling functions of Sutherland \& Dopita \shortcite{sutherland.93}.
For the same value of $T_0$ as before and $Z=0.1$ Z$_{\odot}$ 
the critical points, corresponding to
the same catastrophic transitions, are $b_1=0.001$ and $b_2=0.097$.
The Hopf bifurcation occurs somewhat earlier, at $b_3=0.136$.
In the zero metallicity case 
$b_1=0.000\,48$ and $b_2\approx0.029$, but the nature of the saddle
connection at $b_2$ is different [positive trace case, see
Guckenheimer \& Holmes \shortcite{guckenheimer.83} 
for details], and therefore an unstable limit
cycle appears before the second key bifurcation, at $b\le b_2$. These
details, however, do not change the general picture of the phase
transition due to the structural stability of bifurcations involved.

The main conclusions of this subsection are the following: (i) the
distributed sources and sinks of mass and energy in the hot gaseous halo
may be locally balanced and the thermodynamic equilibrium is stable if
$b>b_{\rmn{crit}}(T_0,\: Z)$; (ii) non-equilibrium solutions settle to the
steady state from any initial conditions in its huge domain of
attraction, and (iii) the equilibrium gas density is proportional to
$\sqrt{\rho_*}$. It is worth emphasizing here that the phenomenon of
phase transition in the system is not due to SN heating, but
is due to a certain combination of the heating and distributed mass
deposition. 

\subsection{The role of feedback} As shown in the previous
subsections, the stable equilibrium can be reached by the system with
no feedback heating. If star formation occurs in the condensed
material, a relatively small energy feedback can modify the dynamics.
This problem will be studied in detail elsewhere. Here we concentrate
only on the effects that feedback heating has on the positions of
the fixed points in the phase plane. It is assumed that the feedback
does not alter the growth rate of condensational instability
({\ref{GrowthRate}). Then equation (\ref{Tem}) must be rewritten as
\begin{equation}
{ \rmn{d} T\over \rmn{d} t}= {1\over\rho}\!\left\{\!\alpha\rho_*(T_0-T)
+\dot\rho_{\rmn{ti}}[T_1-(\gamma-1)T] - \rho^2\Lambda/c_{\rmn{v}}\!\right\},
\label{TemFb}
\end{equation}
to account for the additional heating. The new quantity
$T_1=\varepsilon_{\rmn{fb}}T_{\rmn{fb}}$ is introduced as the feedback
temperature. It is easy to check that near the equilibrium equation 
(\ref{TemFb}) is equivalent to (\ref{Tem}) with $T_0+T_1$ substituted
for $T_0$ in equation (\ref{Tem}). Therefore the presence of additional
feedback heating does not change the structure of equilibria; only the
characteristic heat source temperature increases, providing a
higher equilibrium temperature for the gas. The role of feedback is
somewhat more complex in situations that are far from equilibrium 
due to a time delay of the feedback star formation.

\section{Hydrostatic recycling of isothermal gas} 
Perhaps 90 per cent of the papers on cooling flows contain an 
explanation for the flow's existence as due to a short cooling time-scale
in the central region of a cluster or a group, where a cD or giant
elliptical galaxy is located. It will be
shown in this section that, while this condition is necessary for the
flow to occur, it is not always sufficient. Note that the
model discussed here is based on the same physics that is
incorporated in most of the cooling flow models. The only difference
is that special attention is paid to make the description more
self-consistent and to incorporate the stability analysis in order to
restrict the parameter space and to avoid problems in numerical
simulations\footnote{For example, if non-linear relaxation oscillations
of the thermodynamic state of the gas occur, special care must be taken
to ensure the convergence of the numerical method in use, cf. Fria\c{c}a 
\shortcite{friaca93}.}.
\subsection{Does cooling gas necessarily flow?}
Simple comparison of equations (\ref{GasDenPowerLaw}) and (\ref{GasDen})
demonstrates that a self-consistent hydrostatic model for isothermal
gas recycling in the hot corona of a giant elliptical galaxy can be
constructed if  
\begin{equation}
\beta(\sigma_{*_0},T_0,Z,b)={1\over2}, \label{Key}
\end{equation}
i.e. in the same case in which the $\rho_*$ distribution provides
the observed optical surface brightness profile (see Section
\ref{GalMod}).
The gas metallicity, $Z$, must be uniform over the galaxy.
Equation (\ref{Key}) can be used as an additional constraint to reduce
the dimension of the parameter space of the model. For example, the
definition for $\beta$ (equation \ref{Beta}), equation (\ref{GasTem}), and
the equilibrium condition (\ref{Teq}) allow one to calculate the value of $b$
for given $T_0$, $\sigma_*$, and $Z$ as 
\begin{equation}
 b=\left[\left({k T_0 \over
2\gamma\mu m_{\rmn{p}}\sigma^2_*}-1\right)(2-\nu_{\rmn{eq}})-1\right]^{-1}. 
\label{Bi}
\end{equation}	
Finally, the integration constant $C$ in equation (\ref{GasDenPowerLaw})
can be obtained from equation (\ref{GasDen}):  
\begin{equation}
C=\sqrt{{\alpha(c_0^2-\gamma c^2)\over (\gamma-1)\Lambda(T,Z)}}. \label{IntC}
\end{equation}

The rate of recycling is a strong function of radius and scales as
$t_{\rmn{s}}^{-1}\sim\rho_*^{1/2}\sim\rho$ in the equilibrium. As the
star formation rate in the cooling gas is proportional to the stellar
density in the galaxy, the newly born stars and the associated heating are
distributed in such a way that they do not distort the structure of
existing gaseous and stellar components. Therefore the recycling is
homologous. 

For a comparison with cooling flow models one can formally calculate
the mass deposition (accretion) rate for the recycling model as  
\begin{equation}
\dot M(r) = \int^r_0{4\pi r^2\dot\rho_{\rmn{ti}}(r)\rmn{d}r} = \alpha
M_*(r).\label{Mdot} 
\end{equation}
For $r\gg r_{\rmn{c}}$ one has $\dot M\sim r^{3-n}$, if $\rho_*\sim r^{-n}$,
and, thus the model predicts a scaling $\dot M\propto r$, if the stellar
density distribution is that of the isothermal sphere. 
Generic
slopes of optical surface brightness profiles for giant elliptical
galaxies are in the range from 1.65 to 2.1 \cite{mihalas.81}.
Therefore $-0.1\la\rmn{d}\log\dot M(r)/\rmn{d}\log r\la0.4$. While giant
ellipticals 
have relatively steep surface brightness profiles and
correspondingly flat
$\dot M(r)$, cDs, having flatter brightness distributions, must have
steeper mass deposition profiles. 

If the radiative cooling function $\Lambda(T,Z)$ is assumed to be known,
the list of input parameters of the model includes the following values: 
$\rho_{\rmn{DM}_0}$, $\rho_{*_0}$, $\sigma_*$, $\alpha$, $T_0$, and $Z$. The
model could be applied in several consecutive steps. First, the
ratio $(\rho_{\rmn{DM}_0}+\rho_{_0})/\rho_{*_0}$ and $\rho_{*_0}$ must be
determined as the best-fitting values for the Poisson's equation solutions to
adjust the optical surface brightness profile. Secondly, the temperature and
metallicity, derived from the spectral analysis of the X-ray data,
provide the condensation efficiency $b$ for the assumed value of
$T_0$. An iterative procedure must be applied to get the value
of $T_0$ that satisfies the observed X-ray luminosity. If the
iterations converge, finally, all involved density distributions are
known and the projected distribution of the X-ray emission may be
compared with the observed one.

Evidently, it is not always possible to find a
hydrostatic solution that fits the observed temperature estimates
$kT\ge1.5$ keV, if the only heat source is SN Ia events with a
`standard' rate. If the gas inside the cooling radius is indeed so hot,
there are two possible solutions of this problem: 
either repetitive gravitational heating is important due to 
violent changes in the gravitational potential of the cluster (or
group) in a series of merging events, or the feedback star
formation heated the gas up. To distinguish between these two or to
find the proportions in which both are involved, the chemical evolution of
the gaseous halo must be studied. Simple considerations for the gas
enrichment in the course of hydrostatic recycling are presented in the
next subsection. 

The recycling model represents a special exact steady-state solution of
the general time-dependent hydrodynamic problem  
\begin{equation}
\frac{\partial \rho}{\partial t}+\nabla\cdot(\rho \bmath{v})=\alpha
\rho_* - \dot\rho_{\rmn{ti}}, \label{Mass}
\end{equation}
\begin{equation}
\frac{\partial (\rho \bmath{ v})}{\partial t}+\nabla(\rho
\bmath{v}^2+p)=\rho\nabla\phi - \dot\rho_{\rmn{ti}}
\bmath{v}, \label{CompleteSystem}
\end{equation}
\vspace{-5pt}
\begin{eqnarray}
\lefteqn{\frac{\partial E}{\partial t}+\nabla\cdot[\bmath{v}(E+p)] = } 
\nonumber \\ 
& & \alpha\rho_* e_* -\dot\rho_{\rmn{ti}}(E+p)/\rho-\rho^2\Lambda +\rho
\bmath{v}\cdot\nabla\phi+\nabla\cdot\bmath{q}, \label{Energy}
\end{eqnarray}
in which {\em hydrostatic} and {\em thermal} balance are decoupled.
Here $p=(\gamma-1)e\rho$ is the pressure, $e$ is the specific internal 
energy, $e_*=c_{\rmn{v}} T_0$, the energy density
$E=\rho(e+v^2/2)$, and 
{\boldmath{$q$}} is the heat flux. 
Numerical solution of equations (\ref{Mass})--(\ref{Energy}) can be used 
to check
the convective stability of the gas in the equilibrium recycling model and
to study possible settling solutions in the inflow--outflow context.
This requires multidimensional simulations and will be the subject of a
separate paper. 
It is known from time-dependent numerical experiments on
cooling flows that the distributed mass sink stabilizes the flow against the
cooling catastrohe for a sufficiently high condensation rate 
\cite{meiksin88,friaca93}. The general mechanism of stabilization and the 
domain of the parameter space where it operates have been 
described comprehensively in Section 3 above. Thus, numerical simulations 
of thermally unstable gas flows {\em must} take into account the 
condensational instability.
This, however, modifies the gas dynamics, because the gas becomes inelastic
\cite{kritsuk94}. 
A new type of elementary solution appears, representing a sort of
subsonic condensation wave, slowly propagating in the direction
opposite to the local pressure gradient and `reloading' the flow. The
structure of wave fronts follows the random velocity field in
the gas and therefore must be filamentary. Perhaps systems of bright
filaments, observed in optical emission lines in some `cooling
flows', could be remnants of violent non-uniform condensation
events in the core of the gas distribution during post-merging transitions
to hydrostatic equilibrium. 

\subsection{Coronal gas enrichment} The hot gas metallicity is determined
by the history of stellar  mass loss and supernova activity in the
galaxy. Keeping the notation of Loewenstein \& Mathews 
\shortcite{loewenstein.91}, one can write
an equation for the {\em i}th element mass fraction
$z_i=\rho_i/\rho$ in the ISM:
\begin{equation}
{\rmn{d}z_i\over \rmn{d}t}={\rho_{*} \over\rho}(\alpha_*
y_{*,i}+\alpha_{\rmn{sn}}y_{\rmn{sn,}i}-\alpha z_i), \label{Chem}
\end{equation}
where $y_{*,i}$ and $y_{\rmn{sn,} i}$ are the yields of the element, 
coming from stellar mass loss and SN Ia explosions, respectively. Under
hydrostatic steady-state conditions equation (\ref{Chem}) reduces to
\[ z_{i, \rmn{eq}}={\alpha_* y_{*,i}+\alpha_{\rmn{sn}}
y_{\rmn{sn,}i} \over
\alpha}\approx
y_{*,i}+{\alpha_{\rmn{sn}}\over\alpha}y_{\rmn{sn,} i},\] and
therefore the metallicity of the ISM may reflect that of the stellar
population  with the additional contribution from supernovae. The
time-scale for gas enrichment, 
\begin{equation}
t_z={\rho \over \rho_*}\left[{\alpha_*\left({y_{*,i}\over
z_i}-1\right) +\alpha_{\rmn{sn}}\left({y_{\rmn{sn,}i}\over
z_i}-1\right)}  
\right]^{-1},
\end{equation}
is proportional to the recycling time-scale $t_{\rmn{s}}$, if all values in
square brackets do not depend on the radius. Hence, if at some moment the
hydrostatic gas with uniform $Z$  has lower abundances than the
stellar matter, a negative gradient of metallicity is 
established in the course of evolution. It can be erased later,
if the system has enough time to reach its `chemical' equilibrium.
This would not be the case if the stellar metallicity distribution (and,
therefore, $y_{*,i}$) in the galaxy were non-uniform. There are
indications, based on magnesium indices as tracers of metallicity,
that, while giant elliptical galaxies are inferred to have high
metallicities, their metallicity gradients are low, resulting in
comparatively metal-rich stellar haloes \cite{thomsen.89}.

The simplified picture above does not take into account some important
effects, which can alter the metallicity distribution and the absolute
value of metallicity in the halo of a galaxy. First, if the enriched
gas of stellar winds and supernova shells are not very well mixed into
the hot IGM, but rather cool down after kinetic energy exchange with
the hot medium, then there is no direct relation of the gas
metallicity to the supernova rate or stellar metallicity of the
galaxy. Secondly, the environmental effects could be important for gaseous
haloes of dominant group/cluster members, as substructure
merging in a 
galaxy distribution can heat the gas, and mix the enriched intragalactic
medium with low-metallicity intragroup/cluster gas. While it seems to
be possible to account for the first complication in the model proposed 
[just substituting a cooling function of higher $Z$ in equation 
(\ref{GrowthRate})],
one has to take care in applying the model to the hotter halo of a
dominant member of a potentially dynamically young aggregation of
galaxies, where the gas temperature still keeps a record of the recent
merging event, but not one of the current SN activity. Adjusting the 
thermal equilibrium condition in this case would require an unacceptably
high supernova rate (and/or feedback heating) to get a higher value of
$T_0$. 

\section{Discussion of observational issues}\label{Obs}
It was pointed out in Section \ref{GalMod} that stellar surface brightness
distributions, obtained with $\beta=1/2$, are fitted well by the
$R^{1/4}$-law. The scaling relation of the one-zone model
showed that a
mass deposition (star formation) rate proportional to $\rho^2$ allows
stable recycling of the hot gas over the whole galaxy with zero bulk
motion of the gas. It will be illustrated here that the $\beta=1/2$ model
fits observed optical and X-ray surface brightness
distributions well
and reproduces major correlations. 
\subsection{X-ray surface brightness distributions} It is convenient to
fit observed X-ray surface brightness distributions $\Sigma(R)$ with a
simple formula, which is a generalization of the modified Hubble profile, 
\begin{equation}
\Sigma(R)=\Sigma_0\left[1+\left({R\over
r_{\rmn{c}}}\right)^2\right]^{-3\beta_{\rmn{fit}}+0.5} \label{BetaFit}
\end{equation}
[see, e.g., Binney \& Tremaine \shortcite{binney.87}], where 
$\Sigma_0$, $r_{\rmn{c}}$, and
$\beta_{\rmn{fit}}$ are free 
parameters and $R$ is the distance from the centre in projection. Due
to the different natures of X-ray thermal emission from diffuse gas and
optical stellar light, the model with $\beta=1/2$, which implies
$\rho_*\propto\rho^2$, provides optical and X-ray surface brightness
distributions that match each other\footnote{In this notation the
surface brightness of stellar light follows equation (\ref{BetaFit}) with
$\beta_{\rmn{fit},*}$ instead of $\beta_{\rmn{fit}}$, deprojected spatial
distributions of stellar and gas mass density follow 
$\rho_*=\rho_{*_0}[1+x^2]^{-3\beta_{\rmn{fit},*}}\sim 
x^{-6\beta_{\rmn{fit},*}}$ 
and $\rho=\rho_0[1+x^2]^{-3\beta_{\rmn{fit}}/2}\sim x^{-3\beta_{\rmn{fit}}}$,
respectively, where $x=r/r_{\rmn{c}}$, and $r_{\rmn{c}}$ could be
different for the 
gas and the galaxy according to instrumental characteristics if the real
core is unresolved. Note that $\beta$ in (\ref{GasDenPowerLaw}) and
$\beta_{\rmn{fit}}$ in (\ref{BetaFit}) are different values.}.  
Therefore, $\beta_{\rmn{fit}}\approx\beta_{\rmn{fit, *}}$. 
This is known to be the case, as far as relatively isolated elliptical
galaxies are concerned [see, for example, {\em Einstein} IPC and HRI
data on a sample of six early-type galaxies, which includes three
Virgo cluster members: NGC~4472 (M49), NGC~4636, and NGC~4649 (M60)
\cite{trinchieri..86}; {\em ROSAT} PSPC observations of NGC~4636 
\cite{trinchieri...94}, NGC~4365 and
NGC~4382 (M85) \cite{fabbiano..94}; analysis of optical and X-ray
surface photometry of  NGC~1399, the central cD galaxy in the Fornax
Cluster \cite{killeen.88}]. 

In order to check if this is also true for M87, the X-ray dominant
galaxy in the centre of a reasonably rich cluster, the
results of {\em ROSAT} PSPC and HRI observations were analysed and
compared with 
optical photometry of M87 at faint light levels in the {\em B} and
{\em V} bands. The
optical data are available to distances of $\approx260d_{15}$
kpc\footnote{The distance of M87 is assumed to be 15 Mpc throughout
this paper.}
along the major axis and $\approx90$ kpc on the minor axis
\cite{carter.78}. The azimuthally averaged slope of the optical surface
brightness profile is $-1.66$ and corresponds to $\beta_{\rmn{fit},*}=0.44$,
while $\beta_{\rmn{fit}}$ for the averaged PSPC X-ray image is
$0.45\pm0.01$ (B\"ohringer 1994, private communication). Analysis
of the NW sector of the HRI image, which is not contaminated by the
radio components, gives $\beta_{\rmn{fit}}=0.45\pm0.01$. The comparison
demonstrates a good agreement of X-ray and optical surface brightness
distributions within the cooling radius, and similar orientations of
the major axes of the outer optical and X-ray isocontours of M87. 

The available surface brightness distributions allow one to determine the
profiles of the mass deposition rate. Optical data give
$\beta_{\rmn{fit}}=0.42$ for the
cD galaxy NGC~1399 \cite{thomas....86,killeen.88,grillmair......94},
the cooling radius is known to be $\sim68$ kpc \cite{thomas....86},
and the stellar mass estimate is
$M_*(68$ kpc$)\simeq10^{12}$ M$_{\odot}$ \cite{jones.94}. Therefore,
according to equation (\ref{Mdot}), $\dot M(r)\simeq1.4\,(r/68$
kpc$)^{0.5}$ M$_{\odot}\:$yr$^{-1}$ for the adopted
$\alpha=4.7\times10^{-20}$ s$^{-1}$. The mass deposition estimate in
this case is close to values 2.2--3.0 M$_{\odot}$ yr$^{-1}$, obtained
by Thomas et al. \shortcite{thomas....86} for a cooling flow model.
Similarly, for M87 the cooling radius
and mass estimates are nearly the same: 74 kpc  \cite{stewart...84},
and $M_*(70$ kpc$)\simeq10^{12}$ M$_{\odot}$ \cite{jones.94}.
The mean slope of the optical surface brightness distribution is $-1.66$,
therefore $\dot M(r)\simeq 1.4\,(r/70$
kpc$)^{0.34}$ M$_{\odot}\:$yr$^{-1}$. In this case the
power index and the absolute value of the mass deposition
rate disagree with the estimate, based on the `standard' cooling flow
analysis of {\em Einstein} observations: $\dot M(r)\simeq10\,(r/70$
kpc$)^{2/3} $ M$_{\odot}\:$yr$^{-1}$, for $5\la r\la70$ kpc 
\cite{stewart...84}.
However, within the field of view of the {\em Einstein} Observatory SSS
($\sim3$ arcmin in radius) and FPCS (3 arcmin $\times30$ arcmin)
our result
$\dot M(3$ arcmin $\simeq13.1$ kpc$)\simeq0.8 $ M$_{\odot}\:$yr$^{-1}$ 
is in rough agreement with the upper estimates neglecting heating:
$\dot M_{_{\rmn{SSS}}}\simeq2.3 $ M$_{\odot}\:$yr$^{-1}$ \cite{mushotzky.88} 
and 
$\dot M_{_{\rmn{FPCS}}}\simeq1.6\pm0.3 $ M$_{\odot}\:$yr$^{-1}$ 
\cite{canizares...82}, which have been accordingly rescaled.

\subsection{X-ray gas temperatures} When thermal instabilities occur
in an accretion flow, two different spectral components are
present in thermal X-ray emission, corresponding to terms
$\dot\rho_{\rmn{ti}}e$ (conventionally, softer X-rays from {\em
isobarically} cooling comoving gas) and $\rho^2\Lambda$ (harder
emission from the diffuse gas component) in 
equation (\ref{Energy}).
 The spectral emissivity of such a composition is  
\begin{equation}
\epsilon_{\nu}(r)=\rho^2(r)\Lambda_{\nu}(T)+\dot\rho_{\rmn{ti}}(r)
c_{\rmn{v}}\Gamma_\nu(T), \label{Emiss}
\end{equation}
where
\begin{equation}
\Gamma_{\nu}(T)\equiv{\gamma
\over\gamma-1}\int_0^T{{\Lambda_{\nu}(T)\over\Lambda(T)}}\rmn{d}T
\label{Gam} 
\end{equation}
\cite{rwhite.87b}. If the gas drops out due to thermal
instability in equilibrium with stellar mass loss in the hot
hydrostatic halo of a galaxy, the relative contributions from cooling 
diffuse gas and cooling condensates are fixed by the equilibrium
conditions, and have similar spatial distributions. In this case equation
(\ref{Emiss}) can be transformed into
\begin{equation}
\epsilon_{\nu}(r)\propto 
\rho_*(r)\Lambda_{\nu}(T)\left[1+q{\Lambda\over
T\Lambda_{\nu}}\Gamma_\nu(T)
\right], 
\end{equation}
where for the adopted deposition law 
\begin{equation}
q={b(2-\nu_{\rmn{eq}})\over b+1}. \label{Qu}
\end{equation}
The radiative cooling function in equations (\ref{Emiss}) and (\ref{Gam}) must
take into account non-equilibrium ionization effects for temperatures
less than several $10^6$~K (see, e.g., Sutherland \& Dopita 1993).
\nocite{sutherland.93}

While the idea of distributed mass deposition in cooling flows is
widely exploited in physical models
\cite{thomas86,rwhite.87a,sarazin.89,meiksin88,meiksin90,david..90,friaca93},
analysis of observed spectra is usually based on one- or
two-temperature fits, which appear to be satisfactory, mainly
due to the 
limited spectral resolution of the available detectors. This technique
does not account for the emission of cooling condensates, and
temperature estimates, based on spectral information, suffer from
limited spectral and spatial resolution of the detectors, 
model-dependent deprojection procedures, uncertain abundances (and their 
gradients) in the hot ISM, and unknown H\,{\sc i} column densities.

Emission from condensates
exceeds the diffuse emission at energies below $\sim2$ keV when the
background temperature is $10^8$~K, and below $\sim0.8$ keV when the
background temperature is $10^7$~K \cite{rwhite.87b}. In the case of the
{\em ROSAT} PSPC detector it will make a substantial contribution to
the harder energy band of the detector, due to a blend of iron L-shell
lines, if the gas condensations cool down from a temperature of about
$3-4\times10^7$~K. In contrast, for background temperatures
lower than $\sim10^7$~K, the dominant contribution from {\em isobaric}
cooling will be in the softer energy band. If spectral analysis does
not take into account the emission from cooling condensations, the
resulting pressure (and gravitating mass) estimates could contain
errors of up to a factor of $\sim 2-3$. Consequently, the observational
estimate of  $\beta_{\rmn{spec}}=\sigma_*^2/(kT/\mu m_{\rmn{p}})$ is 
uncertain by the same factor. 

Reanalysis of the {\em Einstein} IPC, HRI, FPCS, and SSS
observations of M87 has demonstrated the disagreement of single-phase
models with mass determinations based on the optical data, while the
available X-ray data are explained well \cite{tsai94a}. The total enclosed
mass, derived from the model at $r<10$ arcmin [i.e. right in
the central `cooling flow' region, where the gas temperature drops 
\cite{boehringer.....94,nulsen.94} to values lower than the
$kT=2.05\pm0.16$ keV determined by {\em Ginga} (Koyama, Takano
\& Tawara 1991)\nocite{koyama..91}], 
falls well below the mass estimate from the optical data. 
The multiphase model adequately explains the X-ray data and predicts total
masses, which agree with optical measurements \cite{tsai94b}. The
central gas temperature for this model is about $1.8\times10^7$~K. It
seems that the main root of the problem is the lack of
self-consistency in the standard procedure implemented for the
gravitating mass determination in the `cooling flow' region, which
is based on the hydrostatic assumption on the one hand and does not
account for the distributed mass deposition (as the only means to
keep the thermally unstable cooling medium hydrostatic) 
on the other hand. The best-fitting multiphase model of Tsai 
\shortcite{tsai94b}
does not show any dramatic decrease of the temperature near the
centre (which is typical for cooling flow models), but shows only a modest
positive temperature gradient, $\rmn{d} \log{T}/\rmn{d} \log{r}=0.114$, 
which is
much smaller than that of the electron number density 
($\rmn{d} \log{n_{\rmn{e}}}/\rmn{d}\log{r}=-0.491$ 
in the core of radius $r_{\rmn{c}}=6.64$ kpc and
$-1.36$ at $r\gg r_{\rmn{c}}$). This can be considered as an argument in 
favour of 
the isothermal approximation, adopted above, and justifies the usage of 
equation
(\ref{IsothHydG}) for gravitating mass estimates. 
\subsection{X-ray gas metallicity} The distribution of iron abundance,
as follows from the {\em Ginga} observation of the Virgo Cluster,
gives a mean value for the central region
within $1^{\circ}$ of M87 of $Z\sim0.5$ Z$_{\odot}$. Because the angular
profile of the abundance distribution is comparable to that of the
{\em Ginga} collimator response ($1^{\circ}$), the detailed structure
remains unresolved. {\em Broad Band X-Ray Telescope} ({\em BBXRT}) 
observations of
the interstellar medium in NGC~1399 indicated a mild positive temperature
gradient ($kT=1.0-1.1$ keV for the central pixel of 4 arcmin in
diameter and $kT=1.1-1.2$ keV for the outer pixels at $\sim7$ arcmin from 
the centre) and a subsolar metallicity with a negative gradient on
the same scale, although the statistical uncertainties do not preclude
a constant metallicity \cite{serlemitsos....93}. 

Spatially resolved
spectra, obtained with the {\em ROSAT} PSPC, can provide information
on possible abundance gradients in the hot gas. Rapidly cooling
regions in the centres of `cooling flows' in NGC~1399 and NGC~4472 show an
apparent trend of decreasing abundance with increasing radial distance
(by a factor of $\sim\!\!2$ for a scale of $\sim\!\!70$ kpc), when $T$, $Z$, 
and
the hydrogen column density, $N_{\rmn{H}}$, are treated as free
parameters for the spectral fitting procedure
\cite{forman.....93,jones.94}. If $N_{\rmn{H}}$
is held fixed, then the gradient is smaller \cite{forman.....93}.
Similarly ambiguous results were obtained by Trinchieri et al.
\shortcite{trinchieri...94} in their detailed study of {\em ROSAT} PSPC data
on NGC~4636. If uniform 100 per cent cosmic abundances were assumed then 
two-temperature components were required to fit the observed
spectra; if 20 per cent abundances were assumed 
then one component provided a fit of the same quality.
New data from the {\em Advanced Satellite for Cosmology and
Astrophysics} 
({\em ASCA}) have confirmed the lower abundances for NGC~4636, however, with 
a mild negative
gradient: from $\sim\!\!0.3$ at $R_{\rmn{e}}\simeq8.4$ kpc to less than
$\sim\!\!0.2$ of the solar value at $5R_{\rmn{e}}$ \cite{mushotzky.....94}.
Also there is good agreement between {\em ROSAT} and {\em ASCA}
temperature profiles, $T\propto r^{\sim0.2}$ over a scale of $\sim30$ kpc
\cite{trinchieri...94}. Even lower metal abundances,
$Z\sim0.15$ Z$_{\odot}$, 
measured with {\em ASCA} were reported for NGC~1404 and NGC~4374 by 
Loewenstein et al. \shortcite{loewenstein.......94}.

Although the observations do not give reliable information on the
metallicity distribution in hot coronae, an assumption of uniform
abundances with solar ratios, adopted above, could
be an appropriate zero approximation for the region inside the
cooling radius. Very low metal abundances detected by {\em ASCA} together
with typical temperatures $kT\sim0.75$ keV would imply
incomplete mixing of gas injected by SNe. However, spectral fits based on
the multiphase gas model are needed to confirm the low abundance estimates.  
\subsection{$L_{\rmn{X}}$--$L_{B}$ correlation} Using simple
geometrical 
arguments, the scaling relation of the one-zone model (\ref{GasDen}),
and the virial 
condition $\rho_{0_*}r^2_{\rmn{c}}=9\sigma^2_*/4\pi G$ [see, e.g.,
Binney \& Tremaine \shortcite{binney.87}], one can write 
$L_{\rmn{X}}\propto \rho_0^2\Lambda r_{\rmn{c}}^3 
\propto\rho_{*_0}c_0^2 r_{\rmn{c}}^3 
\propto\rho_{*_0}\sigma_*^2 r_{\rmn{c}}^3 
\propto\sigma_*^4 r_{\rmn{c}}$. For galaxies on the
fundamental plane $L_{B}\propto \sigma^{1.3}r_{\rmn{c}}$
\cite{djorgovski.90}, 
therefore 
$L_{\rmn{X}}\propto L_{B}^{3.1}r_{\rmn{c}}^{-2.1}$. Using the
Faber--Jackson law 
for bright ellipticals [$\sigma_*\propto L_{B}^{0.25}$ 
\cite{faber.76a}] and the above relationship for the
fundamental plane, it is easy to eliminate $r_{\rmn{c}}$ and get 
$L_{\rmn{X}}\propto L_{B}^{1.7}$.
Note that the Faber--Jackson relation can be
derived from a simple self-regulated star formation scenario for
a massive ($M\ge10^{11}$ M$_{\odot}$) spheroidal system \cite{lin.92}.
This scenario implies the initial expulsion of gas, combined with that
resulting from later stellar evolution, which may form a source of
gas, feeding the hot haloes of individual galaxies. 
In agreement with this, observations show that the relation
depends on the galaxy luminosity.  While for the brightest
($M_{B}\la-20$) ellipticals
$L_{B}\propto\sigma_*^{4.2\pm0.9}$, for the less 
luminous ones ($M_{B}\ga-20$) the deviations from the
Faber--Jackson law may be considerable: $L_{B}\propto\sigma_*^{2.4\pm0.9}$
\cite{davies....83}. Thus, one has 
$L_{\rmn{X}}\propto L_{B}^{1.65}$ and 
$L_{\rmn{X}}\propto L_{B}^{2.15}$ 
for bright and faint ellipticals, respectively.
These results can be compared with  
$L_{\rmn{X}}\propto L_{B}^{1.7\pm0.3}$ 
(Canizares, Fabbiano \& Trinchieri 1987)\nocite{canizares..87} 
and $L_{\rmn{X}}\propto L_{B}^{2.18\pm0.20\\}$
(Donnelly, Faber \& O'Connell 1990)\nocite{donnelly..90}. 
Note that the derived relation of X-ray to
optical luminosities is quite different from the naive scaling
$L_{\rmn{X}}\propto L_{B}$ for SN-heating dominated models. 

In the framework of the recycling model the deviations from the `mean'
X-ray to optical luminosity relationship [the observed rms residuals of
$\log{L_{\rmn{X}}}$ are $0.33-0.35$ \cite{donnelly..90}] may be
attributed to the combined effects of:
(i) galaxy-to-galaxy differences in the spectra of
perturbations, which stimulate thermal instabilities in the hot gas;
(ii) different rates of SN activity and/or stellar mass loss rates;
(iii) variations of the chemical composition of the hot gas.
Each of these circumstances determines the equilibrium gas content of
a galaxy and most of them may be environmentally influenced (see also 
Section 5.6 below).
The most dramatic deviations may occur if the efficiency of condensation $b$
becomes lower than its critical value $b_{\rmn{crit}}$. This could 
stimulate oscillations 
of the gas density with amplitude $\sim\!\! 5$ \cite{kritsuk92} and, 
therefore, 
initiate the random walk of a galaxy about the mean value of $L_{\rmn{X}}$. The
possibility of such destabilization is one of the natural explanations for 
considerable scatter in the $\log{L_{\rmn{X}}}-\log{L_B}$ plane.

\subsection{$L_{\rmn{X}}$--$T$ correlation} For the $\beta$-model
$T\propto\sigma_*^2$ and therefore, using the $L_{\rmn{X}}-L_{B}$
relation above and the 
Faber--Jackson law, one obtains $L_{\rmn{X}}\propto T^{3.4}$. For the
high- and low-luminosity ellipticals $L_{\rmn{X}}\propto T^{3.44}$ and
$L_{\rmn{X}}\propto T^{2.54}$, respectively. 
Note that this scaling relation is similar to
$L_{\rmn{X}}\propto T^{3.3}$, for a sample of galaxy clusters observed by
{\em Ginga} \cite{arnaud94}. For cluster cooling flows the 
observed X-ray luminosity--temperature relation has the same scaling,
but depends also on the mass deposition rate (estimated 
with the `standard' cooling flow analysis technique):  $L_{\rmn{X}}\propto
T^{3.3}\dot M^{0.4}$ \cite{fabian...94a}. 

\subsection{Gas content and environmental effects}
The relation of the SN-associated heat source temperature $T_0$ to the
equilibrium gas temperature and density, as well as $Z$ and $b$, is
illustrated here in Table 1, computed using equations
(\ref{GasTem}), (\ref{Teq}) and (\ref{IntC}) and the cooling
functions of Sutherland \& Dopita \shortcite{sutherland.93}.
It shows that higher metallicities result in lower $T_0$
values and lower gas densities (X-ray luminosities) for a fixed
temperature of the hot diffuse component of the ISM. Also the higher
condensation efficiency provides a lower $L_{\rmn{X}}$. When the
efficient feedback star formation is triggered by the recent merging
event, a higher equilibrium X-ray luminosity is expected.
This might be the generic case for dominant galaxies in rich and,
therefore, more active environments. In relatively isolated galaxies,
as is illustrated in the following, the standard value of the SN rate can
readily explain the observed gas temperatures. Note, however, that the
uncertainties of the $T_0$ estimates inherit the uncertainties of the model 
cooling
functions and their first derivatives. Therefore the values of $T_0$ may be 
only 30-50 per cent accurate.  

\begin{table}
\caption{Thermodynamic equilibria for hot coronal gas.}
\begin{center}
\begin{tabular}{lcccccc}
\multicolumn{7}{c}{$Z=$ Z$_{\odot}$}\\ \hline\hline
 & \multicolumn{2}{c}{$b=0.5$} &\multicolumn{2}{c}{$b=0.7$} &
\multicolumn{2}{c}{$b=0.9$}\\ \cline{2-3}\cline{4-5}\cline{6-7} 
$T_{0_7}$&$T_7$&$C_{-15}$&$T_7$&$C_{-15}$&$T_7$&$C_{-15}$\\
2.0&0.54&3.34&0.57&3.30&0.60&3.25\\
2.5&0.61&3.95&0.65&3.83&0.70&3.71\\
3.0&0.69&4.38&0.76&4.20&0.82&4.05\\
3.5&0.77&4.73&0.99&4.26&1.16&4.03\\
4.0&1.06&4.71&1.27&4.57&1.35&4.51\\
4.5&1.29&5.13&1.40&5.08&1.47&5.01\\
5.0&1.40&5.63&1.52&5.55&1.60&5.46\\
5.5&1.50&6.10&1.64&5.97&1.75&5.83\\
6.0&1.60&6.52&1.79&6.33&1.89&6.21\\
7.0&1.85&7.26&2.02&7.10&2.12&6.97\\
8.0&2.04&7.99&2.21&7.79&2.36&7.57\\
9.0&2.21&8.65&2.47&8.30&2.62&8.09\\
10.&2.43&9.16&2.68&8.84&2.83&8.63\\ \hline
\end{tabular}
\medskip

\begin{tabular}{lcccccc}
\multicolumn{7}{c}{$Z=0.32$ Z$_{\odot}$}\\ \hline\hline
 & \multicolumn{2}{c}{$b=0.5$} &\multicolumn{2}{c}{$b=0.7$} &
\multicolumn{2}{c}{$b=0.9$}\\ \cline{2-3}\cline{4-5}\cline{6-7} 
$T_{0_7}$&$T_7$&$C_{-15}$&$T_7$&$C_{-15}$&$T_7$&$C_{-15}$\\
2.0&0.52&4.52&0.57&4.41&0.60&4.32\\
2.5&0.61&5.25&0.66&5.10&0.70&4.97\\
3.0&0.69&5.86&0.76&5.60&0.85&5.30\\
3.5&0.79&6.28&0.94&5.84&1.02&5.64\\
4.0&0.95&6.53&1.13&6.15&1.27&5.88\\
4.5&1.13&6.83&1.31&6.56&1.37&6.47\\
5.0&1.30&7.24&1.40&7.13&1.51&6.87\\
5.5&1.38&7.78&1.56&7.43&1.64&7.27\\
6.0&1.50&8.18&1.66&7.88&1.73&7.77\\
7.0&1.69&8.98&1.84&8.71&2.02&8.30\\
8.0&1.87&9.71&2.12&9.21&2.25&8.98\\
9.0&2.11&10.2&2.33&9.81&2.50&9.48\\
10.&2.29&10.8&2.55&10.3&2.68&10.1\\ \hline
\end{tabular}
\end{center}
\medskip
{\small The heat source temperature $T_{0_7}$ and
the equilibrium gas temperature $T_7$ are given in units of $10^7$~K.
The integration constant $C_{-15}$ in equation (\ref{GasDenPowerLaw}) is
computed for the stellar mass loss rate $\alpha=4.7\times10^{-20}$
s$^{-1}$ and given in units of $10^{-15}$ (g cm$^{-3}$)$^{1\over2}$.}

\end{table}

Recent {\em ASCA} observations of bright elliptical galaxies in the
Virgo Cluster revealed hard X-ray components with colour temperature
$kT\ge2$ keV in addition to extended thermal X-ray emission
of temperature $kT\sim1$ keV \cite{matsushita.15.94}. This hard
component is primarily
attributed to the integrated emission from discrete X-ray sources,
associated with the optical galaxy. For galaxies located in the area
of extended cluster X-ray emission, the discrete sources alone cannot
account for all the hard X-ray flux detected with {\em ASCA}. It was
assumed, therefore, that the hard components of NGC~4406 (M86) and
NGC~4374 (M84) are contaminated 
by the foreground/background Virgo intracluster medium emission. In
the case of NGC~4472 the hard-band image is elongated and displaced from 
the optical galaxy and from the soft-band image as well. This is
interpreted as the locally enhanced ICM emission plus the
discrete-source contribution, although their respective contributions remain
unspecified. The temperature of the soft emission from NGC~4472,
$kT=0.83\pm0.04$ keV, corresponds to an SN heating temperature
$T_0\approx3.4\times10^7$~K for the observed metallicity
$Z\sim0.42$ Z$_{\odot}$ and the condensation efficiency $b=0.7$.
For NGC~4636 located at the outskirts of the Virgo Cluster, {\em ROSAT} PSPC
data imply a similar hard ($kT>0.9$ keV) component, which is more extended than
the soft one ($kT\simeq0.6-0.8$ keV) and probably is not directly related to
the galaxy. However, this is only true if 100 per cent cosmic abundances 
are assumed. 
With 20 per cent cosmic abundances only one temperature component 
($kT\simeq0.6-1.0$ 
keV) is required for spectral fits \cite{trinchieri...94}.

The {\em BBXRT} data for NGC~1399, a cD galaxy in a poor cluster,
suggest a gas temperature $kT=1.0-1.2$ keV and metallicities of about
$0.5$ Z$_{\odot}$ with large uncertainties \cite{serlemitsos....93}. The
analysis of two-temperature models revealed (similarly to the
Virgo galaxies) a hard
component of X-ray emission with $kT\sim6$ keV and provided a lower
temperature $\sim1$ keV and a higher metallicity $0.76$ Z$_{\odot}$ for the
primary soft component. This would imply a characteristic heat source
temperature $T_0\approx4\times10^7$~K.  

An accurate deprojection analysis of the {\em ROSAT} PSPC spectral
data on M87 provided
temperature estimates for the centre of the `cooling flow' region
($r\le3.33$ arcmin $\simeq15$ kpc) of $kT=1.1-1.4$ keV \cite{nulsen.94}.
Such a temperature would require $T_0\approx5\times10^7$~K for the
subsolar metallicity $Z\sim0.5$ Z$_{\odot}$ and an even lower value for
solar metallicity. The condensation efficiency $b=0.7$ roughly corresponds to
$q=0.874$ -- the value for the best-fitting multiphase model of Tsai
\shortcite{tsai94b}, see equation (\ref{Qu}). Note that the temperature
estimate was made in Nulsen \& B\"ohringer
\shortcite{nulsen.94} for the region 
where the errors due to contamination by the ICM emission are small.
The straightforward approach would require $T_0\sim9\times10^7$~K for
the gas temperatures detected with wide-beam instruments [e.g.
$kT\sim2$ keV, {\em Ginga} \cite{koyama..91}]. Note, however, that the
multiphase model based on {\em Einstein} data gives a central
temperature estimate of $T\simeq1.8\times10^7$~K.

Although in all three cases discussed above the low gas temperature
estimates were obtained on the basis of a single-phase description, if
they are considered as rough estimates for the temperature of the diffuse 
component of the cooling gas, the standard SN Ia rate appears to be
sufficient for thermodynamic equilibrium 
at the measured temperatures. The analysis of {\em ASCA}
observations, based on the multiphase model, would help to check whether
the diffuse component of the multiphase ISM in central dominant
galaxies actually has a low temperature $\la1$ keV and nearly solar
metallicity. In this case a $kT\la1$ keV galactic atmosphere might be
embedded in the extended intracluster gas with a higher temperature and lower
metallicity. {\em ASCA} data on NGC~4696/Centaurus lend support to 
this option \cite{fukazawa.......94}.
Such a complex quasi-hydrostatic configuration can be described
with a hybrid model, including the recycling interior of the dominant
galaxy matched at the cooling radius [where
$t_{\rmn{c}}(r_{\rmn{cool}})=H_0^{-1}$] with an 
isothermal $\beta$-model for the hotter tenuous intracluster gas. If
the ratio of the stellar-to-gas mass is low and the cooling radius is
larger than the extent of the optical galaxy, then a cooling flow is
unavoidable. It starts from the cooling radius, enters the galaxy, and
disappears, providing material for star formation.

\section{Conclusions}

A purely hydrostatic model for the hot gas in normal elliptical galaxies 
has been presented as an alternative to the `standard' cooling flow picture.
The model independently develops an option suggested by Thomas et al.
\shortcite{thomas....86} that perhaps there are no radial gas flows, so that
all gas present at some radius originated there (e.g. from stellar mass
loss) and cools and is deposited {\em in situ}. Energy injection from SNe and
feedback heating due to star formation from deposited material have been
also incorporated in the description of equilibrium gas recycling. 
Formally, the model is quite similar to the steady-state cooling flow 
formulation 
of Sarazin \& Ashe \shortcite{sarazin.89} but with zero 
bulk motion of the gas. 
(Note that the notion of the cooling flow itself implies
highly subsonic velocities and there are no observational data that would
directly confirm the existence of the {\em flow}.) Therefore, the recycling 
model, representing a `minimal cooling flow', gives the exact lower estimates 
of the mass deposition rate in elliptical galaxies and fills the gap in a
family of models for hot gaseous coronae, being a fully self-consistent
hydrostatic model which for the first time takes into account the whole set 
of sources and sinks of mass and energy. It has the advantage
of an analytical 
description of gas equilibria, which allows the study of their linear 
stability. 
However, the global thermal-convective stability of such a hydrostatic 
recycling 
atmosphere of a galaxy must be the subject of a separate numerical 
investigation. 

Summarizing the results of this study, it is worthwhile to stress that
the galaxy, located in the centre of the cooling region, may play
an important role, providing a complex mechanism for hydrostatic
support of thermally unstable gas. The moderate amount of SN
Ia heating in combination with the distributed mass sink is able to
compensate the radiative cooling and the stellar mass loss in
such a way that quasi-isothermal hot gas is kept in both {\em
hydrostatic} and {\em thermodynamic} equilibrium. 
The latter becomes possible due to a first-order phase transition, 
which separates the stable hot gas phase for a sufficiently high efficiency of 
condensation. Having no energy input and/or mass sink it is impossible to
avoid a cooling catastrophe in the centre of the `cooling flow' region,
where radiative cooling and stellar mass loss are efficient.

\section*{Acknowledgments}

This work has been supported in part by the Russian
Foundation of Fundamental Research (project code 93-02-02957) and by a
grant of the American Astronomical Society. The author is grateful to
Andi Burkert, Makoto Hattori, and Ewald M\"uller for helpful comments on
the manuscript and to the staff of MPA and MPE for the warm hospitality.

\bsp

\label{lastpage}

\end{document}